\documentclass[superscriptaddress,reprint,amsmath,amssymb,aps,pra]{revtex4-1}
 
\usepackage{mathtools,amsmath}
\usepackage{amssymb}
\usepackage{amsfonts}
\usepackage{bm}
\usepackage{dsfont}
\usepackage{graphicx}
\usepackage[utf8]{inputenc}
\usepackage[T1]{fontenc}
\usepackage[english]{babel}
\usepackage{xcolor}
\usepackage{tikz}
\usepackage{tikz-feynman}
\usepackage{ulem}

\graphicspath{{../figures/}}

\newcommand{\p}{\partial}

\newcommand{\bigexp}[1]{\exp{\left\{ #1 \right\}}}

\DeclareMathOperator{\acosh}{acosh}

\begin{document}
\title{Multiphoton resonance in a driven Kerr oscillator in presence of high--order nonlinearities}
\date{\today}
\author{Evgeny V. Anikin}
\affiliation{Skolkovo Institute of Science and Technology, 121205 Moscow, Russia}
\author{Natalya S. Maslova} 
\affiliation{
    Quantum Technology Centrum, Department of Physics, 
    Lomonosov Moscow State University, 119991, Moscow, Russia
    }
\author{Nikolay A. Gippius}
\affiliation{Skolkovo Institute of Science and Technology, 121205 Moscow, Russia}
\author{Igor M. Sokolov}
\affiliation{Institut für Physik and IRIS Adlershof, Humboldt Universität 
             zu Berlin, Newtonstraße 15, 12489 Berlin, Germany}
\begin{abstract}
We considered the multiphoton resonance in the periodically driven quantum oscillator with Kerr nonlinearity 
in the presence of weak high–order nonlinearities. Multiphoton resonance leads 
to the emergence of peaks and dips in the dependence of the stationary occupations of the stable states on detuning. We demonstrated that due to
high–order nonlinearities, these peaks and dips acquire additional fine structure and split into several closely spaced ones. 
Quasiclassically, multiphoton resonance is treated as tunneling between the regions of the oscillator phase portrait, and the fine structure of 
the multiphoton resonance is a consequence of a special quasienergy dependence of the tunneling rate between different regions of the classical phase portrait. 
For different values of damping and high--order nonlinearity coefficients, we identified the domain of quasienergies where 
tunneling strongly influences the system kinetics. The corresponding tunneling term in 
the Fokker--Planck equation in quasienergy space was derived directly from the quantum master equation. 
%


\end{abstract}
\maketitle
\section{Introduction}
For decades, bistable and multistable systems attract researchers' attention in many areas of physics. Bi-- and multistability 
has been observed in many experimental setups including nonlinear--optical systems \cite{Azadpour2019}, 
lasers \cite{Li2017a}, nanomechanical systems \cite{Pistolesi2018}, optical cavities interacting with ultracold atoms \cite{Gothe2019} 
or magnonic systems \cite{Wang2018}. Recently, it became possible to observe bistability in systems operating with only a few excitation quanta
\cite{Wang2019} \cite{Winkel2020} \cite{Muppalla2018}. Such systems are promising candidates for the generation of squeezed states which are important for 
decreasing the noise--signal ratio in quantum measurements \cite{PhysRevB.100.035307}.
Moreover, they can be useful for the creation of entangled states which are crucial for applications in 
quantum information processing and safe quantum communications systems. 


There exists a class of bistable systems that can be modeled as a nonlinear oscillator mode with Kerr nonlinearity driven by external resonant 
or parametric excitation. Such models describe a wide range of physical systems including the Fabry--Perot microcavities with nonlinear filling \cite{Gibbs1976}, 
whispering gallery resonators, laser systems near threshold \cite{Bonifacio1978}, polariton microcavities, superconducting nonlinear resonators 
\cite{Wang2019} \cite{Winkel2020} \cite{Muppalla2018}. 
On the classical level, the model of a driven nonlinear oscillator has two stable 
stationary states with different field amplitudes. With account for thermal noise, transitions between these states become
possible. As the states 1 and 2 have different field amplitudes and intensities, they can be distinguished experimentally, 
for example, via a cross--Kerr induced shift in some probe mode. In the experiment, it is possible to observe random switching between the stable states 
\cite{Muppalla2018}. Thus, it is of high interest to calculate the occupation probabilities of the stable states and the transition 
rates between them.

At small or moderate numbers of quanta circulating in the mode, quantum effects become important. Interestingly, when the number of quanta in the mode 
is several dozens, the quantum effects can be treated within the quasiclassical approximation, and it is still possible to use the 
classical concepts of the classical phase portrait and stable states. One of the most pronounced quantum effects 
is related with tunneling between different regions of the phase portrait of the classical oscillator. Tunneling transitions modify the occupation probabilities of the 
classical stable states and the transition rates, namely, they increase the occupation of the high--amplitude stable state and thus lead to enhanced 
excitation of the mode \cite{Maslova2019} \cite{Anikin2019}. In fact, tunneling between different regions of the phase portrait corresponds to the 
quasiclassical treatment of multiphoton transitions, namely, the excitation of the oscillator with simultaneous absorption of many external field quanta. 
A similar relation between multiphoton transitions and tunneling is known in the theory of multiphoton ionization of atoms \cite{Keldysh1965}.

In the model of a single oscillator mode with Kerr nonlinearity, tunneling and multiphoton transitions are especially important when the 
resonance condition is fulfilled. If no higher nonlinearities are present, this occurs when the detuning 
between the driving field and the oscillator mode is an integer or half--integer multiple of the Kerr frequency shift per quantum. This property 
follows from a special symmetry of the model Hamiltonian \cite{Anikin2019}, and because of this, the eigenstates of the quantum Hamiltonian correspond to 
superpositions of quasiclassical states belonging to different regions of the phase portrait.
Because of that, the dependence of the higher--amplitude and lower--amplitude states populations on detuning has 
pronounced peaks and drops at integer and half--integer detuning--nonlinearity ratio. However, in real systems, small higher--order nonlinearities 
always exist together with Kerr nonlinearity. It is of high interest to find out how their presence modifies the structure of multiphoton resonance. 

In this manuscript, we consider the model of a quantum driven nonlinear oscillator which includes high--order nonlinearities as small 
corrections. Together with numerical simulations, we utilize the analytical approach of the Fokker--Planck equation in the quasienergy space 
with tunneling term obtained from the full quantum master equation. We demonstrate that in presence of high--order nonlinearities, the 
multiphoton resonance peaks in the 
occupations of the high--amplitude stable state split into several smaller ones with different widths and amplitudes. The magnitude of the splitting turns out to
be proportional to high--order nonlinearity coefficients. In addition, we extend the analysis of previous works \cite{Maslova2019} \cite{Anikin2019}
to the case of finite damping having the order or being larger than tunneling and multiphoton splitting between the Hamiltonian eigenstates from different 
regions of the phase portrait. 
\section{The model of a quantum driven nonlinear oscillator}
\label{sec:model}
We consider the model of a bistable driven system consisting of a resonant mode with Kerr--like 
nonlinearity \cite{Drummond1980}, \cite{Risken1987} and additional higher--order nonlinearities. 
The effective Hamiltonian of the system in the rotating--wave approximation reads 
\begin{equation}
    \begin{gathered}
    \label{eq:quant_bist_ham}
        \hat{H} =-\Delta \hat{a}^\dagger \hat{a} + 
        \frac{\alpha}{2}(\hat{a}^\dagger \hat{a})^2 + \hat{V} + f(\hat{a} + \hat{a}^\dagger),\\
        \hat{V} = \sum_{q=3}^{\infty}\alpha_q(a^\dagger a)^q
    \end{gathered}
\end{equation}
The eigenstates of this effective Hamiltonian are the approximations of the exact Floquet states of the full time--dependent Hamiltonian, 
and the eigenvalues give the Floquet quasienergies.
The parameter $\Delta$ is the detuning between the driving field and the resonant oscillator frequency, $\alpha$ is the Kerr coefficient, $\alpha_q$ 
is the $2q$--order nonlinearity coefficient, and $f$ is proportional to the amplitude of the driving field. In the following, we 
will mostly focus on the case of six--order nonlinearity, $q = 3$. 

The statistical properties of this model with account for weak interaction with the dissipative environment should be studied 
using the quantum master equation (QME):
\cite{Haken1965}, \cite{Risken1965},
\cite{Graham1970}, \cite{Drummond1980}, \cite{Risken1987}:
\begin{equation}
    \label{eq:lindblad_eq}
    \p_t \rho = i[\rho, \hat{H}] + \frac{\gamma}{2}
                    \left(2\hat{a}\rho \hat{a}^\dagger - \rho a^\dagger a\right.\\ - a^\dagger a \rho 
                                     \left. + 2N[[a,\rho],a^\dagger]\right),
\end{equation}
where $\gamma$ is the coupling strength with the dissipative environment, 
and $N$ is the number of thermal photons at the external field frequency.
\begin{figure}[h]
    \includegraphics[width=\linewidth]{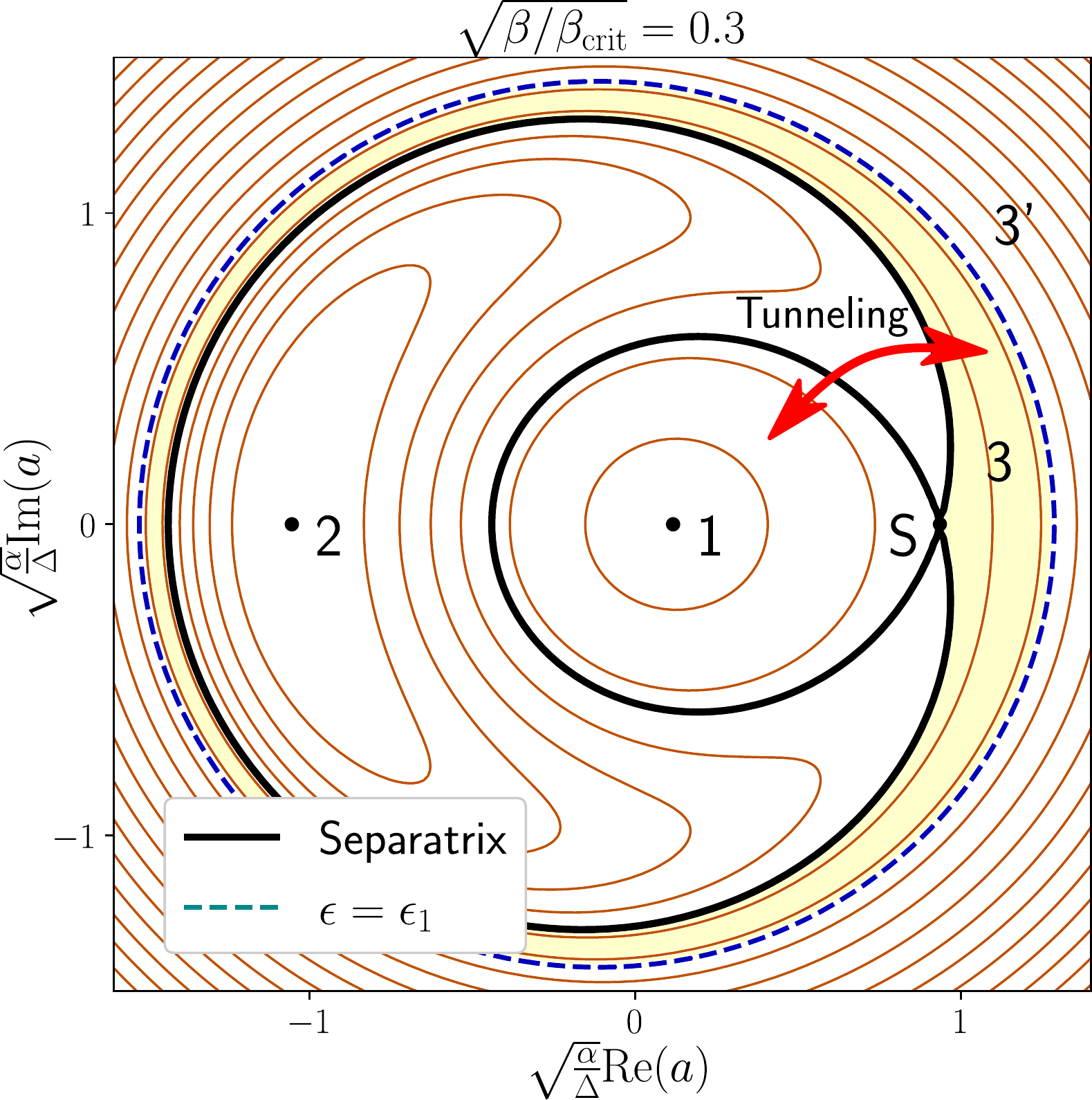}
    \caption{The classical phase portrait of the nonlinear oscillator with the
             Hamiltonian \eqref{eq:quant_bist_ham} for $f/f_\mathrm{crit} = 0.3$, $\alpha_3 = 0$. The blue dashed line denotes a classical 
            trajectory in region 3 having the same quasienergy $\epsilon_1$ as the stable state 1. From region 1, the system can exhibit 
            tunneling transitions to the subregion of region 3 enclosed by the separatrix and this trajectory.
             }
    \label{fig:phase_portrait}
\end{figure}
Both unitary dynamics governed by the system Hamiltonian \eqref{eq:quant_bist_ham} and dissipative dynamics described by QME \eqref{eq:lindblad_eq}
can be treated quasiclassically, if the Kerr nonlinearity is sufficiently small, $\Delta \gg \alpha$. 
While the exact unitary dynamics of the system are described by Heisenberg equations for operators $\hat{a}, \hat{a}^\dagger$, 
one should replace these operators with the c--number field amplitudes $a$, $a^*$ in the system
Hamiltonian \eqref{eq:quant_bist_ham} to obtain the classical limit. The time evolution of the classical field amplitudes $a$ and $a^*$ is the motion
along the classical trajectories given by the contour lines of the classical Hamiltonian $H(a,a^*)$. Also, according to Bohr--Sommerfeld rule, 
the eigenstates of the quantum Hamiltonian correspond to a discrete set of trajectories on the classical phase portrait 
in the quasiclassical limit. Importantly, the Bohr--Sommerfeld description
does not take into account quantum tunneling which will be discussed below.
For the dissipative dynamics in the same limit, the QME can be 
transformed into the classical 2D \cite{Vogel1988} \cite{Maslova2007} or 
1D Fokker--Planck equation \cite{Vogel1990} \cite{Maslova2019}, which 
is equivalent to classical Langevin equations containing the Hamiltonian 
term, the damping term and 
the noise term. The quasiclassical approach demonstrates good agreement with the full quantum simulations even at moderate numbers of photons ($\sim 20$)
\cite{Maslova2019} circulating in the mode.

A prominent feature of the classical phase portrait
is bistability, which is present for field values not exceeding the critical value $f_\mathrm{crit} = \sqrt{4\Delta^3/27\alpha}$ (at $\hat{V} = 0$). 
In this case, there are two stable stationary states 1 and 2.
In addition, there exists an unstable stationary state S 
and a self--intersecting trajectory (separatrix) passing through S. The separatrix divides the phase portrait into 
regions 1 and 2 containing the 
corresponding stable stationary states and the outer region 3 
(see Fig.~\ref{fig:phase_portrait}). The classical trajectories from
region $2$ have quasienergies $\epsilon$ such as 
$\epsilon_2 < \epsilon < \epsilon_{sep}$, where $\epsilon_r$ is the
quasienergy of the classical stable state $r = 1,2$, 
and $\epsilon_{sep}$ is the quasienergy of the unstable stationary state. 
For the trajectories from region 1,
$\epsilon_{sep} < \epsilon < \epsilon_1$, and for the trajectories from 
region 3, $\epsilon > \epsilon_{sep}$. For additional details on the role of different quasienergy domains, please
see Fig.~2 in Ref.~\cite{Anikin2019}.
Also, the presence of small 
higher--order nonlinearities doesn't change the qualitative structure of the classical phase portrait.

According to both the quasiclassical treatment of the model using the quasiclassical Fokker--Planck equation (FPE) \cite{Vogel1990} and the
full quantum treatment based on QME \cite{Drummond1980} \cite{Risken1987}, 
the system persists in the vicinity of the classical stable states 1 and 2 most
of the time. Also, rare noise--induced transitions 
between the stable states occur. Thus, the probabilities to find the system close to the stable states 1 and 2, $\bm{P}_1$ and $\bm{P}_2$, can be identified with
the probabilities to find the system in regions 1 and 2 of the
classical phase portrait. In the classical limit, they can be found from 
the stationary solutions of the FPE as the integrals of the probability 
density over the corresponding domain of quasienergies. Beyond
the applicability of FPE, they can be obtained from the stationary 
solutions of the QME.

\section{Tunneling between the regions of the classical phase portrait}
\label{sec:tunneling}
\begin{figure}[h]
     \includegraphics[width=\linewidth]{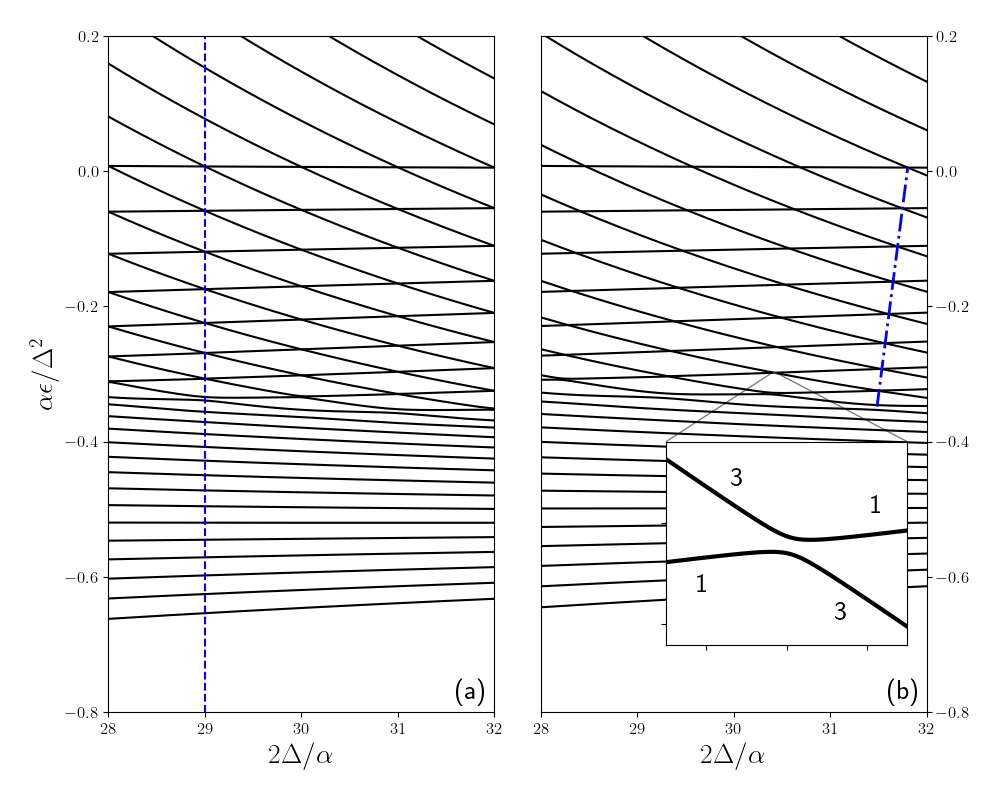}
    \caption{The eigenvalues of the Hamiltonian \eqref{eq:quant_bist_ham} obtained via exact numerical diagonalization
    are shown  for different ratios between the detuning $\Delta$ and nonlinearity $\alpha$ 
    at $f/f_\mathrm{crit} = 0.1$ and (a) $\alpha_3 = 0$, (b)
    $\alpha_3/\alpha = 0.005$. In absence of high--order nonlinearity,
    all anticrossings occur at integer values of $m$ and lie on a single
    vertical line (see the blue dashed vertical line in (a)). 
    This is not the case in presence of six--order 
    nonlinearity, when the anticrossings of 
    quasienergy levels occur at different values of $2\Delta/\alpha$. 
    On the inset, zoomed region of anticrossing between the levels from
    classical regions 1 and 3 is shown.}
    \label{fig:6_ord_spectrum}
\end{figure}

For each classical trajectory in region 1, there exists a trajectory 
with the same value of quasienergy in region 3 (see 
Fig.~\ref{fig:phase_portrait}). 
Quantum mechanics allow the system to undergo a tunneling transition 
between two such classical trajectories, so the Bohr--Sommerfeld 
quasiclassical description of the eigenstates of the quantum Hamiltonian
should be modified with account for tunneling. Actually, the real 
Hamiltonian eigenstates can be considered as quantum superpositions of the
trajectories belonging to different regions of the phase portrait.
However, the tunneling amplitude is exponentially small in comparison 
with the spacing between the quasienergy levels within each region. Because 
of that, the trajectories form superpositions only when a certain resonance
condition for the system parameters is fulfilled. In absence 
of high--order nonlinearities, it 
was shown \cite{Anikin2019} that this happens when the detuning $\Delta$ 
is an integer or half--integer multiple of $\alpha$ independently of 
$f$. This manifests as the anticrossings
of the Hamiltonian quasienergy levels dependence on $\Delta$ at the constant 
driving field. (see the inset on Fig.~\ref{fig:6_ord_spectrum}). Moreover,
a prominent feature of the model without high--order nonlinearities is 
that the anticrossings of many pairs of levels occur simultaneously.
This is a consequence of a special symmetry of the system Hamiltonian, 
namely, the symmetry of the perturbation theory series for the system 
quasienergies $\epsilon_n$ in $f$. Also, it can be seen from the results of 
numerical diagonalization, which are shown in Fig.~\ref{fig:6_ord_spectrum}.

Since the true eigenstates of the Hamiltonian can 
be superpositions of trajectories from regions 1 and 3, it is convenient 
to use the basis of states which are not the eigenstates of the 
quantum Hamiltonian but correspond to a discrete set of 
classical trajectories lying entirely in one of the regions of 
the phase portrait. In such a basis, Hamiltonian is not diagonal, 
and matrix elements corresponding to tunneling transitions between different 
regions of the classical phase space are present. Also, when 
$2\Delta/\alpha$ is close to an integer, the quasienergy levels
group into pairs with very close values of quasienergy, and the tunneling 
matrix element can be retained only between the states within each pair.
Thus, the Hamiltonian in the suggested basis reads
\begin{multline}
    \label{eq:regions_basis}
    \hat{H} = \sum_n \begin{pmatrix}
                        |n,1\rangle & |n,3\rangle
                     \end{pmatrix}
                     \begin{pmatrix}
                        \epsilon_{n1} & t_n\\
                        t_n & \epsilon_{n3}
                     \end{pmatrix}
                    \begin{pmatrix}
                        |n,1\rangle\\
                        |n,3\rangle
                    \end{pmatrix}\\
        +\sum_n \epsilon_{n2} |n,2\rangle\langle n,2|
        +\sum_n \epsilon_{n3'} |n,3'\rangle\langle n,3'|
\end{multline}
Here $|n,2\rangle$ are the states from region 2, and $|n,3'\rangle$ are the 
states from region 3 with quasienergies higher than the states from region 1. 
These states are not affected by tunneling. Then, the states
$|n,1\rangle$ and $|n,3\rangle$ form the pairs of the basis states
from regions 1 and 3 with close values of mean quasienergy $\epsilon_{n1}$ 
and $\epsilon_{n3}$. It is necessary to take the amplitude of tunneling $t_n$
between them, which can be estimated as \cite{Maslova2019}
\begin{equation}
    \begin{gathered}
        \label{eq:tunneling_amplitude}
        t_n \sim \Delta e^{-S_\mathrm{tunn}(\epsilon_n)},\\
        S_\mathrm{tunn} = \frac{\Delta}{\alpha}\int_{q_1}^{q_2} 
        \acosh\left\{ \frac{\frac{\alpha\epsilon}{\Delta^2} + 
            \frac{s^2}{2} - \frac{s^4}{8}}{s\sqrt{2\alpha f^2/\Delta^3}}
        \right\} s\,ds,
    \end{gathered}
\end{equation}
In the integral in the expression for tunneling amplitude, $q_1$ and $q_2$ 
are two branching points of the $\acosh$ function.
 
The anticrossings of the quasienergy levels affect the statistical and 
kinetic properties of the model because of enhanced tunneling 
between the regions of the phase space. It was shown \cite{Maslova2019} 
that tunneling decreases the population of the 
stable state 1 and increases the population of the stable state 2 due 
to the presence of an additional escape 
channel from classical region 1. Thus, 
each anticrossing decreases the population of the stable state 1 
and increases the population of the stable state 2 and the field intensity 
in the mode.

In presence of nonvanishing $\hat{V}$, the anticrossings of 
different pairs of quasienergy levels occur at close but different
values of detuning (see Fig.~\ref{fig:6_ord_spectrum}b). This 
can be explained by considering $\hat{V}$ as a small perturbation. It is 
convenient in the basis introduced above because the averages of $\hat{V}$ over
the basis states can be calculated as the c--function averages over classical 
phase trajectories.

Let us consider a pair of levels $n_1$ and $n_3$ which exhibit
anticrossing at the detuning value 
$\Delta_0 = m_0\alpha/2$, $m_0 \in \mathbb{Z}$, when $\hat{V} = 0$. This means
that $\epsilon_{n1} = \epsilon_{n3}$ at this value of $\Delta$.
When high--order nonlinearities are present, $\epsilon_{n1}$ and 
$\epsilon_{n3}$ acquire first--order corrections, and
the anticrossing of the levels occurs at some 
$\Delta = \Delta_0 + \delta\Delta$. 
By treating $\delta\Delta$ as a perturbation together with $\hat{V}$, 
one can get the expression for the quasienergy differences
\begin{multline}
    \label{eq:energy_differences}
    \epsilon_{n1} - \epsilon_{n3} =
        \delta\Delta
        \left((a^\dagger a)^{11}_{nn} - (a^\dagger a)^{33}_{nn}\right)
        + V^{11}_{nn} - V^{33}_{nn},
\end{multline}
where $\langle n,r|\hat{O}|n',r'\rangle \equiv O^{rr'}_{nn'}$ for any
operator $\hat{O}$.  The new anticrossing position follows from the equality 
$\epsilon_{n1}(\Delta_0 + \delta\Delta_n, \alpha_q) 
= \epsilon_{n3}(\Delta_0 + \delta\Delta_n, \alpha_q)$:
\begin{equation}
    \label{eq:deltaDelta}
    \delta\Delta_n = \frac{V_{nn}^{33} - V_{nn}^{11}}
    {(\hat{a}^\dagger\hat{a})_{nn}^{33}-(\hat{a}^\dagger\hat{a})_{nn}^{11}}
\end{equation}
When the shifts of the anticrossing positions are considerably smaller 
than $\alpha/2$, the anticrossings are located near
the integer values of $2\Delta/\alpha$. 
The number of anticrossings near each integer $m = \frac{2\Delta}{\alpha}$, 
is proportional to $m$, and their offsets from integer values
are of order $\alpha_3m^2$. Basing on an accurate analysis of the 
quantum master equation, we will show below that level anticrossings give rise
to a set of peaks near integer values of $2\Delta/\alpha$ in the high--amplitude 
stable state occupation.

\section{Multiphoton resonance and the populations of the stationary states}
To analyze the effect of tunneling on the stationary density matrix and the
populations of the classical stable stationary states, 
let us consider the master equation \eqref{eq:lindblad_eq} in 
the basis of states $|n,1\rangle$, $|n,2\rangle$, $|n,3\rangle$ and 
$|n,3'\rangle$ introduced in Section~\ref{sec:tunneling} (see also Eq.~\ref{eq:regions_basis}).  By employing the diagonal approximation in this basis 
and performing the gradient expansion, the classical FPE 
in quasienergy representation can be obtained in the limit of large
$2\Delta/\alpha$, constant ratio $\alpha(N+1/2)/\Delta$ and small $\gamma/\Delta$ \cite{Maslova2019}. 
Tunneling between the regions of the phase portrait is mediated by the nondiagonal elements of the density
matrix and lies beyond this approximation. 
However, it is possible to retain only the density matrix elements
$\rho_{nn'}^{rr'}$ with $n = n'$ (denoted hereafter as $\rho_{n}^{rr'}$), 
because $\rho_{nn'}^{rr'}$ are proportional to $\gamma/(\epsilon_{n1} - \epsilon_{n'3})$ and can be neglected as long as $\gamma$ is small in comparison to 
the quasienergy spacing within each region of the phase portrait.
Also, the matrix elements of the annihilation operator $\hat{a}$ between the states lying in different regions of the phase portrait, $r\ne r'$, 
are exponentially small and can be neglected. Under such approximations, 
the master equation for regions 1 and 3 takes the form
\begin{widetext}
\begin{multline}
\label{eq:lindblad_reduction}
\p_t \rho_{n}^{rr} = \pm it_n(\rho_{n}^{13}-\rho_{n}^{31})
  -\gamma(N+1)\left((a^\dagger a)_{nn}^{rr}\rho_{n}^{rr}
      -\sum_{n'}a_{nn'}^{rr}(a_{nn'}^{rr})^*\rho_{n'}^{rr} \right) \\
    -\gamma N\left(
        (a a^\dagger)_{nn}^{rr}\rho_{n}^{rr} 
        -\sum_{n'}(a_{n'n}^{rr})^*a_{n'n}^{rr}\rho_{n'}^{rr}
        \right), \quad r = 1,3
    \end{multline} 
    \begin{multline}
    \label{eq:lindblad_reduction_nondiag}
    \p_t \rho_{n}^{13} = -i(\epsilon_{n1} - \epsilon_{n3})\rho_{n}^{13} +it_n(\rho_{n}^{11} - \rho_{n}^{33}) \\
      - \frac{\gamma}{2}(N+1)\left( 
          (a^\dagger a)_{nn}^{11} \rho_{n}^{13} + (a^\dagger a)_{nn}^{33} \rho_{n}^{13}
          - \sum_{n'}2a_{nn'}^{11}(a_{nn'}^{33})^* \rho_{n'}^{13}
          \right)\\
        - \frac{\gamma N}{2}\left( 
            (aa^\dagger)_{nn}^{11} \rho_{n}^{13} + (aa^\dagger)_{nn}^{33} \rho_{n}^{13}
            - \sum_{n'}2(a_{n'n}^{11})^*a_{n'n}^{33} \rho_{n'}^{13}
            \right)
        \end{multline}
\end{widetext}
The system of equations \eqref{eq:lindblad_reduction} and  
\eqref{eq:lindblad_reduction_nondiag}
can be transformed 
into continuous form by
considering the density matrix 
elements $\rho_{n}^{11}$, $\rho_{n}^{33}$ as 
continuous functions $P_1$, $P_3$ of $n$
and performing the gradient expansion like in \cite{Maslova2019}. 
For our purposes, it is more convenient to use the quasienergy 
$\epsilon(n) = (\epsilon_{n1} + \epsilon_{n3})/2$ as an independent
continuous variable.  Also, in the stationary case, 
it is possible to express the nondiagonal density matrix elements 
$\rho_{n}^{13}$ from \eqref{eq:lindblad_reduction_nondiag} and to substitute
them into \eqref{eq:lindblad_reduction}. The resulting equations for 
$P_r(\epsilon)$, $r = 1,3$ in the domain of quasienergies
$\epsilon_\mathrm{sep} < \epsilon < \epsilon_1$ read
\begin{multline}
\label{eq:energy_fokker_planck_w_tunn}
\frac{1}{T_r(\epsilon)} \frac{\p}{\p \epsilon}
\left[\gamma K_r P_r + 
 QD_r \frac{\p P_r}{\p \epsilon}\right] 
  \pm \lambda_T(P_3 - P_1) = 0
\end{multline}
where $T_r(\epsilon)$ is the period of motion along the 
classical trajectories, $K_r(\epsilon)$, $D_r(\epsilon)$ are the drift and 
diffusion coefficients in quasienergy space in each region 
of the classical phase 
portrait, $Q = \gamma(N+1/2)$ is the noise intensity, and
$\lambda_T(\epsilon)$ is a coefficient which can be interpreted as the rate of
tunneling transitions between the regions of the classical phase portrait. 
The term with $\lambda_T(\epsilon)$ arises because of 
the presence of nondiagonal
elements of the density matrix, and
the particular form of $\lambda_T(\epsilon)$ will be derived 
directly from the master equations \eqref{eq:lindblad_reduction} and 
\eqref{eq:lindblad_reduction_nondiag}. 
It turns out that $\lambda_T(\epsilon)$ has a 
nontrivial dependence on $\epsilon$, $\Delta$, the 
coefficients $\alpha_q$ in $\hat{V}$ and $\gamma$. Below, we will show that 
the tunneling term $\lambda_T(P_3 - P_1)$ strongly changes the stationary distribution function
and the populations of the stationary states. By examining 
$\lambda_T(\epsilon)$, it is possible 
to explain the structure of resonant peaks in the occupation of the classical stationary state 2
in presence of high--order nonlinearities and finite damping.

\begin{figure}[h]
  \includegraphics[width=\linewidth]{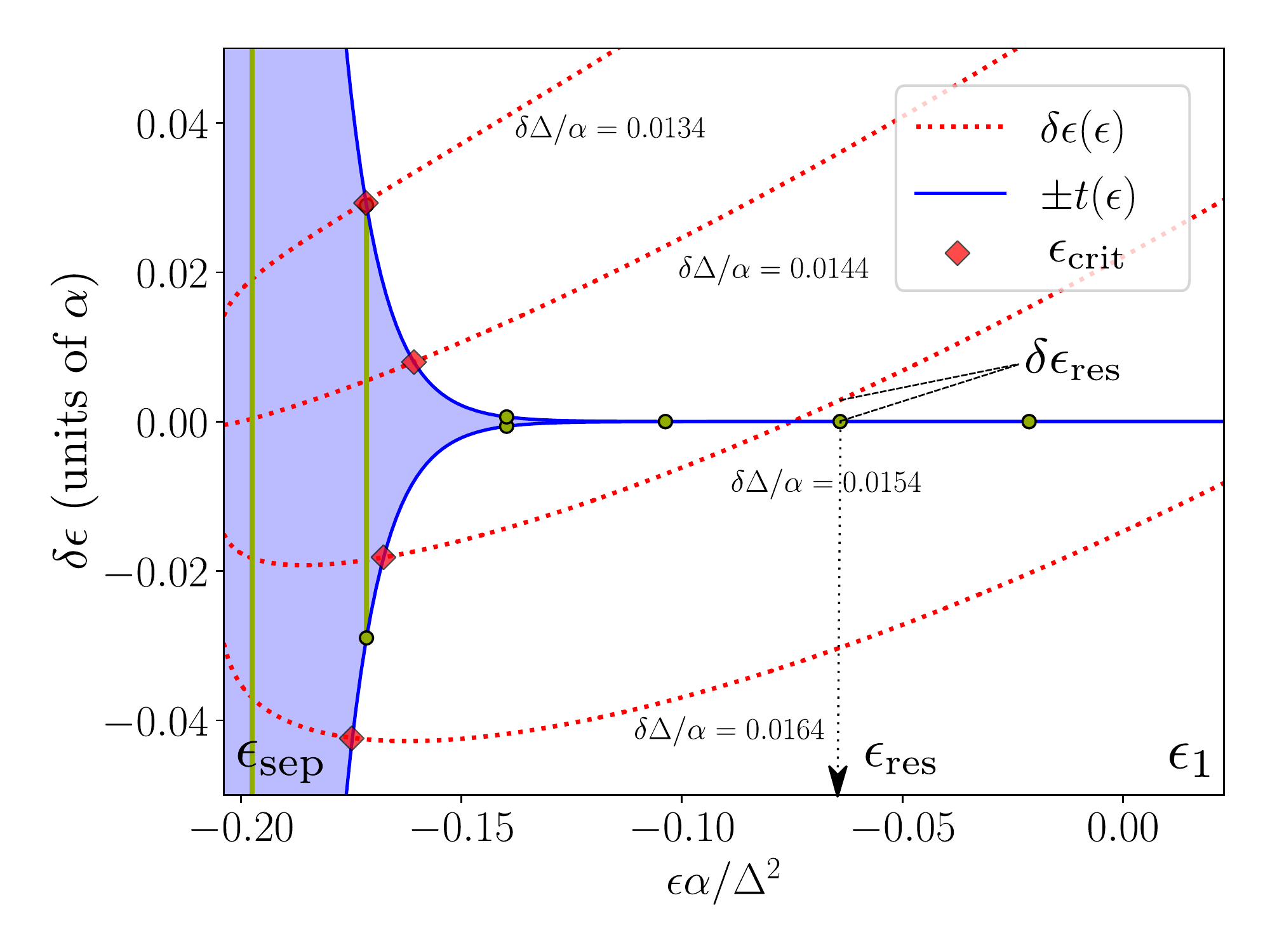}
  \caption{For different detunings $\delta\Delta$, the behavior of $\epsilon_\mathrm{crit}$ 
    and $\epsilon_\mathrm{res}$ is demonstrated by comparing two sides of the 
    inequality \eqref{tunnel_inequality}. The red dashed lines depict $\delta\epsilon_{13}(\epsilon)$ 
    and blue solid lines depict the tunneling amplitude $t(\epsilon)$ for $\alpha_3/\alpha = 10^{-5}$.
  }

    \label{fig:tunnel_regions}
\end{figure}

Before proceeding to the derivation of $\lambda_T(\epsilon)$, let us 
give the qualitative analysis of the role of tunneling between different 
pairs of almost--degenerate states $|n,1\rangle$ and $|n,3\rangle$.
Tunneling between these states has different importance for different $n$:
when 
          \begin{equation}
          \label{tunnel_inequality}
          t_n \gtrsim |\epsilon_{n1} - \epsilon_{n3}|,
          \end{equation}
tunneling is strong and leads to the hybridization of the states $|n,1\rangle$ and $|n,3\rangle$. 
In the opposite case, $t_n \ll |\epsilon_{n1} - \epsilon_{n3}|$, tunneling
can be neglected. The inequality $t_n \gg |\epsilon_{n1} - \epsilon_{n3}|$
holds in two different cases. First, it is always satisfied 
for such $n$ that  $t_n \gg \delta\Delta, \alpha_3$ because 
$|\epsilon_{n1} - \epsilon_{n3}|$ is of order $\delta\Delta$, $\alpha_3$, 
see Eq.~\eqref{eq:energy_differences}. There can be many pairs of states 
$|n,1\rangle$ and $|n,3\rangle$ for which
$t_n \gg \delta\Delta, \alpha_3$. 
Because of the exponential decay of $t_n$ away from 
the separatrix, they lie in the domain of quasienergies
$\epsilon_\mathrm{sep} < \epsilon < \epsilon_\mathrm{crit}$, where 
$\epsilon_\mathrm{crit}$ is is a new parameter depending on $\delta\Delta$, 
$\alpha_3$ and $\gamma$ which we call critical quasienergy. From 
Eq.~\ref{tunnel_inequality}, it follows that $\epsilon_\mathrm{crit}$ is a 
minimal value among the roots of the two equations 
$\delta\epsilon_{13}(\epsilon) = \pm t(\epsilon)$,
where $t(\epsilon)$ and $\delta_{13}(\epsilon)$ are the continuous limits 
of $t_n$ and $\epsilon_{n1}-\epsilon_{n3}$ taken as functions of 
the quasienergy $\epsilon$ (see Fig.~\ref{fig:tunnel_regions}).
Second, even in the case 
$t_n \ll \delta\Delta, \alpha_3$, the inequality \eqref{tunnel_inequality}
still can be satisfied for a single pair of the states
$|n,1\rangle$ and $|n,3\rangle$ for some $n=n_\mathrm{res}$
if $\epsilon_{n1} - \epsilon_{n3}$ passes near zero
at $n_\mathrm{res}$. This is possible because two terms in 
Eq.~\eqref{eq:energy_differences} may have different signs, and physically this
can be interpreted as resonant tunneling through a single pair of 
almost--degenerate states. However, such a pair of states exists only 
when higher--order nonlinearities are present. 

The steps to derive \eqref{eq:energy_fokker_planck_w_tunn} 
from \eqref{eq:lindblad_reduction} and \eqref{eq:lindblad_reduction_nondiag} 
assuming that $\p_t\hat{\rho} = 0$ are as follows.
First, one should express the nondiagonal density matrix elements $\rho^{13}_{n}$ and $\rho^{31}_{n}$ through $\rho^{11}_n$ and $\rho^{33}_n$ using 
Eq.~\eqref{eq:lindblad_reduction}. Then, they should be substituted 
into \eqref{eq:lindblad_reduction}, and the continuous limit should 
be obtained. After the calculation presented in Appendix 
\ref{app:qme_grad_exp}, 
one gets the tunneling rate as 
          \begin{equation}
          \label{eq:lambda_T}
          \lambda_T(\epsilon) = 
          \begin{cases}
          \displaystyle
          \frac{\gamma_{13}(\epsilon)t^2(\epsilon)}
{\delta\epsilon_{13}(\epsilon)^2 + 
  \frac{\gamma_{13}^2(\epsilon)}{4}},
  & \epsilon_\mathrm{sep} < \epsilon < \epsilon_\mathrm{crit}
  \\
    \displaystyle
    \frac{\tilde\gamma^{13}_{n_\mathrm{res}}t^2(\epsilon_\mathrm{res})}
{\delta\epsilon_{13\mathrm{res}}^2 + 
  \frac{(\tilde\gamma^{13}_{n_\mathrm{res}})^2}{4}}
  \frac{\delta(\epsilon-\epsilon_\mathrm{res})}{T(\epsilon)},
  &
  \epsilon_\mathrm{crit} < \epsilon < \epsilon_1\\
    \end{cases},
  \end{equation}
where $t(\epsilon)$ and $\delta_{13}(\epsilon)$ are the continuous limits 
of $t_n$ and $\epsilon_{n1}-\epsilon_{n3}$ taken as functions of 
the quasienergy $\epsilon$, and 
  \begin{equation}
  \gamma_{n13} = \gamma\left((a^\dagger a)_{nn}^{11} + 
      (a^\dagger a)_{nn}^{33} -
      2\sum_{n'} a_{nn'}^{11}(a^{33}_{nn'})^* \right)
  \end{equation}
  \begin{equation}
  \tilde\gamma_{n13} = 
  \frac{\gamma}{2}\left((a^\dagger a)_{nn}^{11} + 
      (a^\dagger a)_{nn}^{33} - 2a_{nn}^{11}(a_{nn}^{33})^*\right)
  \end{equation}
The delta--function term in \eqref{eq:lambda_T} exists only when higher--order 
nonlinearities are present. Below, we will show that it leads to 
emergence of the fine structure of the multiphoton resonance 
peak in $\bm{P}_2$, namely,
several additional narrow side peaks.

After we described the behavior of $\lambda_T(\epsilon)$, let us analyze 
the stationary distribution $P_r(\epsilon)$ over quasienergies.
In each of the domains $\epsilon_\mathrm{sep} < \epsilon < \epsilon_\mathrm{crit}$, 
$\epsilon_\mathrm{crit} < \epsilon < \epsilon_\mathrm{res}$, 
$\epsilon_\mathrm{res} < \epsilon < \epsilon_{1}$, different 
analytical expressions for the stationary distribution function 
can be obtained. 
Due to strong tunneling in the domain  
$\epsilon_\mathrm{sep} < \epsilon < \epsilon_\mathrm{crit}$, the probability
distributions in regions 1 and 3 become almost equal,
$P_1 \approx P_3$. 
By considering the sum of the equations \eqref{eq:energy_fokker_planck_w_tunn}
for $P_1$ and $P_3$, one can obtain a single first--order differential equation for distribution functions $P_{1,3}$. The details of the calculation are 
given in the Appendix \ref{app:qme_grad_exp}. 
The resulting distribution function in the domain $\epsilon_\mathrm{sep} < \epsilon < \epsilon_\mathrm{crit}$ 
turns out to decay exponentially away from the separatrix. This in contrast with the case of the purely classical oscillator, 
for which the distribution function $P_1(\epsilon)$ grows exponentially away from the separatrix.
In the domain $\epsilon_\mathrm{crit} < \epsilon < \epsilon_1$, 
tunneling transitions occur only for the quasienergy 
$\epsilon \approx \epsilon_\mathrm{res}$ due to a delta--function peak in
$\lambda_T(\epsilon)$. Because of that, the stationary distributions in the 
domain $\epsilon_\mathrm{crit} < \epsilon < \epsilon_\mathrm{res}$ are the 
solutions of \eqref{eq:energy_fokker_planck_w_tunn} with 
nonzero probability flow. The flow can be obtained from 
boundary conditions at $\epsilon_\mathrm{res}$ 
obtained by integrating \eqref{eq:energy_fokker_planck_w_tunn} in the vicinity 
of $\epsilon_{res}$. Finally, for $\epsilon > \epsilon_\mathrm{res}$, the 
stationary distributions in regions 1 and 3 coincide with the solutions 
of \eqref{eq:energy_fokker_planck_w_tunn} without tunneling term and with 
zero probability flow. The example of such an analytical solution of 
the FPE is shown in Fig.~\ref{fig:one_channel}, and the detailed calculation
is given in Appendix \ref{app:fpe_stat}.

The solution of the FPE demonstrates that tunneling through the domain 
$\epsilon_\mathrm{sep} < \epsilon < \epsilon_\mathrm{crit}$ and through
the resonant pair increase the population of the stable state 2. 
Now let us analyze how the obtained solutions depend on $\Delta$,
$\gamma$ and $\alpha_3$. 

Let us analyze the behavior of $\epsilon_\mathrm{crit}$. Analyzing its definition
as the minimal root of $\delta\epsilon_{13}(\epsilon) = \pm t(\epsilon)$, 
one deduces that $\epsilon_\mathrm{crit}$ has a sharp peak 
at some value of $\delta\Delta$ and decays to $\epsilon_\mathrm{sep}$ 
away from it. The more $\epsilon_\mathrm{crit}$ is, the more pairs of states 
$|n,1\rangle$ and $|n,3\rangle$ with quasienergies 
$\epsilon_\mathrm{sep} < \epsilon_n < \epsilon_\mathrm{crit}$ 
strongly contribute to tunneling between the regions of the classical 
phase portrait. Thus, the maximum in $\epsilon_\mathrm{crit}$ results in a peak 
of the probability $\bm{P}_2$ to find the system in the classical region 2. 

In addition, the delta--function term 
in $\lambda_T(\epsilon)$ is present when the
condition \eqref{tunnel_inequality} is satisfied for $n_\mathrm{res}$ such as
$t_{n_\mathrm{res}} \ll \delta\Delta,\alpha_3$. 
As discussed in section~\ref{sec:tunneling},
the condition $\epsilon_{n_\mathrm{res}1} = \epsilon_{n_\mathrm{res}3}$ is the condition
of level anticrossing which is satisfied at 
$\delta\Delta = \delta\Delta_{n_{res}}$ defined by Eq.~\eqref{eq:deltaDelta}.
So, at each $\delta\Delta = \delta\Delta_{n}$ being
much larger than the tunneling amplitude $t_{n_\mathrm{res}}$, 
there is also a narrow peak in $\bm{P}_2$. 

Thus, the peaks in the population of the high--amplitude 
stable state dependence on $\Delta$ acquire fine structure 
due to high--order nonlinearity. Namely, a sequence of narrow side 
peaks with the spacing of order $\alpha_3\Delta/\alpha$ arise near the main
resonance, and the number of these peaks is $\sim \Delta/\alpha$. This qualitative 
picture holds until the width of the whole sequence of peaks ($\sim \alpha_3\Delta^2\alpha^2$)
becomes comparable with $\alpha$ and different sequences of peaks start to overlap.


These predictions are 
in good correspondence with the results of numerical solution
of the full quantum master equation \eqref{eq:lindblad_eq}, 
see Fig.~\ref{fig:p1_p2_ratio_w_energies} 
and Fig.~\ref{fig:p1_p2_ratio_alpha_3}.
In Fig.~\ref{fig:p1_p2_ratio_w_energies}, $\bm{P}_2$ 
is shown as a function of $\Delta$
together with the differences of the 
quasienergies of the Hamiltonian eigenstates 
which exhibit anticrossings. Each peak in the probability 
$\bm{P}_2$ of the stable state 2 occupation is located at the value 
of $\Delta$ corresponding to a minimal difference between the 
eigenstates quasienergies.

   \begin{figure}[h]
   \centering
   \includegraphics[width=\linewidth]{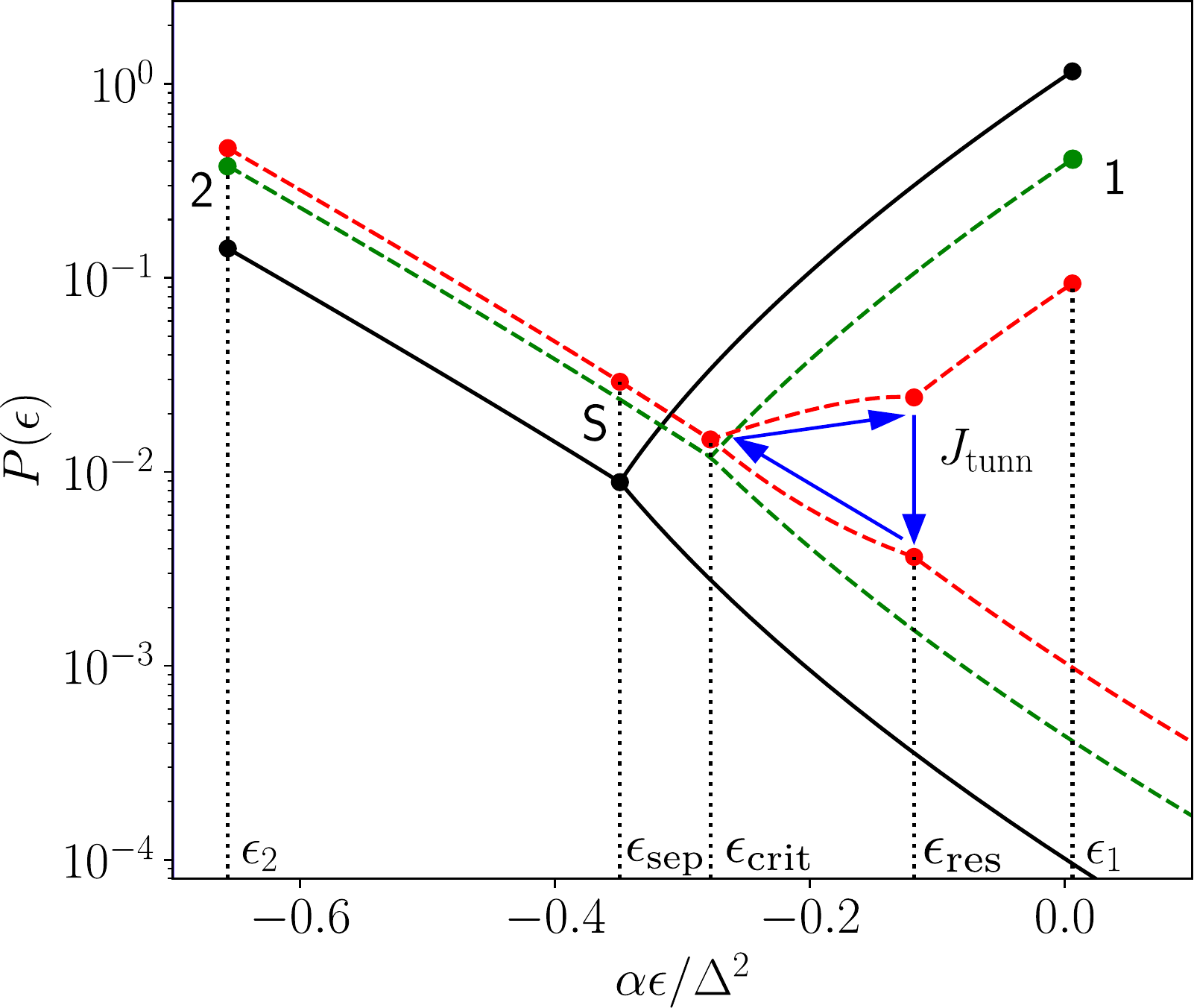}
   \caption{For $f/f_\mathrm{crit} = 0.2$, $\alpha Q/(\Delta \gamma) = 0.1$,
       the theoretically predicted stationary 
       probability distribution function of a nonlinear
       oscillator with six--order nonlinearity near multiphoton 
       resonance is shown by a red dashed line when 
       the tunneling domain 
       $\epsilon_\mathrm{sep} < \epsilon < \epsilon_\mathrm{crit}$ 
       exists together with a 
       single pair of degenerate states at $\epsilon=\epsilon_\mathrm{res}$. 
       For comparison, the probability distribution function is also 
       shown by a green line when the resonant pair of states does not exist, and
       the distribution function of ththee purely classical oscillator is shown by a black solid line. 
   }
\label{fig:one_channel}
\end{figure}

    \begin{figure}[h!]
    \includegraphics[width=\linewidth]{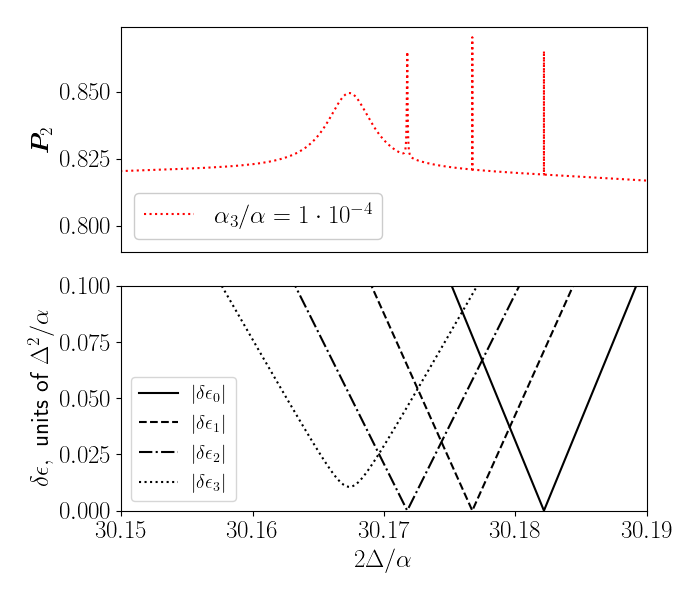}
    \caption{The dependence of the occupation of the higher--amplitude stable state of the quantum driven nonlinear oscillator 
      with six--order nonlinearity on $2\Delta/\alpha$ is shown in (a) for $f/f_\mathrm{crit} = 0.4$, 
           $\alpha_3/\alpha^2 = 1\cdot10^{-4}$. In (b), the differences between the pairs of anticrossing quasienergy levels are shown.
Each peak in $\bm{P}_2(\Delta)$ corresponds to an anticrossing of two quasienergy levels.
    }
\label{fig:p1_p2_ratio_w_energies}
\end{figure}

\begin{figure}
\includegraphics[width=\linewidth]{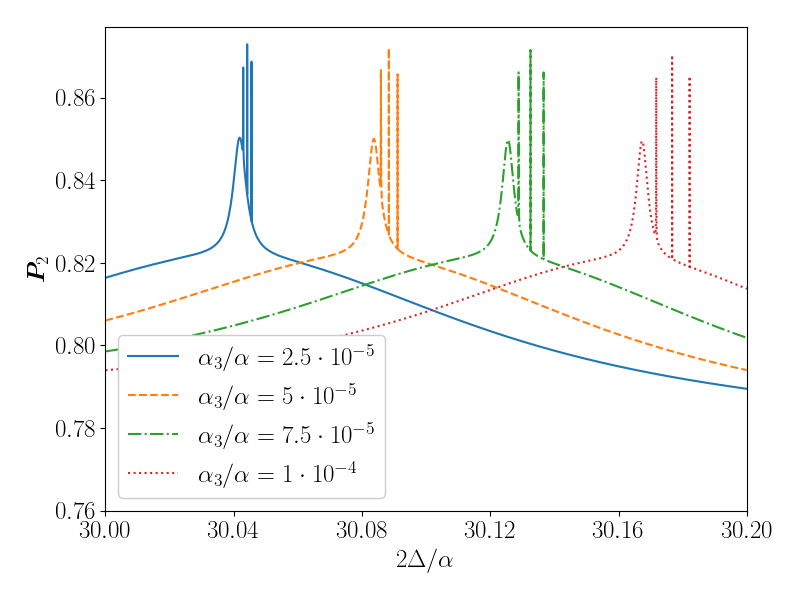}
\caption{For the quantum driven 
nonlinear oscillator with six--order nonlinearity, the dependence of the
probability $\bm{P}_2$ to find the system in the classical region 2
on $\Delta$ is shown in the limit of $\gamma \to 0$ for 
  $f/f_\mathrm{crit} = 0.4$, $N = 3$, and for different values 
of $\alpha_3/\alpha$. The position of each of 
  the peaks corresponding to multiphoton resonance depends linearly on $\alpha_3/\alpha$, and at $\alpha_3/\alpha = 0$ the peaks merge.}
\label{fig:p1_p2_ratio_alpha_3}
\end{figure}

In addition, let us analyze the effect 
of finite damping on the described fine structure of the multiphoton resonance. This can be done simply by analyzing the equations
\eqref{eq:lambda_T} because they are derived from the master equation \eqref{eq:lindblad_reduction}, \eqref{eq:lindblad_reduction_nondiag} which already 
accounts for the effect of damping and the nondiagonal elements of the density matrix.
First, the role of the delta--peak in \eqref{eq:lambda_T} corresponding to 
the resonance between the $n_\mathrm{res}$--th pair of levels depends on the
ratio between $t_{n_\mathrm{res}}$ and the corresponding decay constant $\tilde\gamma_{n13}$. 
Thus, at increasing $\gamma$, the side peaks disappear in the order of increasing $t(\epsilon_n)$.  No side peaks are left when $\gamma$ 
reaches the value of $t(\epsilon_\mathrm{crit}^{\mathrm{max}})$, where $\epsilon_\mathrm{crit}^{\mathrm{max}}$ is the maximum value
of $\epsilon_\mathrm{crit}$ depending on $\delta\Delta$. At larger $\gamma$, the depth of the main peak also becomes $\gamma$--dependent 
because $\lambda_T(\epsilon)$ can be neglected for the quasienergies $t(\epsilon) \ll \gamma$.





\begin{figure}[h]
\includegraphics[width=\linewidth]{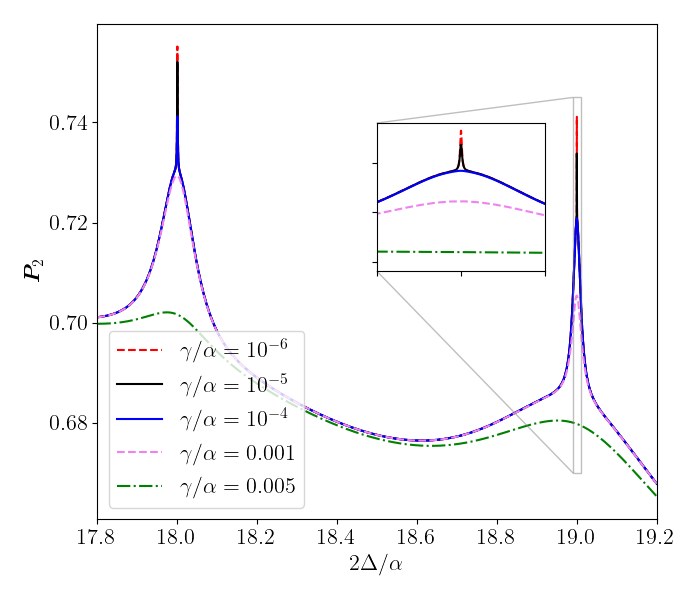}
\caption{For the quantum driven nonlinear oscillator with $\alpha_3 = 0$,
the probability to be in the classical region 2 is shown as a function of 
$2\Delta/\alpha$ for 
different values of $\gamma$. At $\gamma=0$, there are sharp peaks 
at integer  corresponding to multiphoton resonance. At increasing $\gamma$, the drops become smoother.}
\label{fig:varying_gamma}
\end{figure}

\section{Conclusions}
In conclusion, we analyzed the effect of multiphoton resonance on the populations of the stable states of the quantum nonlinear oscillator in resonant
driving field. By including the tunneling term into one--dimensional quasiclassical Fokker--Planck equation in quasienergy space, we demonstrated that the 
mean--field intensity exhibits peaks near the values of external field frequency corresponding to multiphoton resonance. These peaks were associated 
with the anticrossings of the quasienergy levels of the oscillator. Also, we considered the effect of the higher--order nonlinearities on the structure of these 
peaks. We showed that due to high--order nonlinearities, the intensity peaks corresponding to multiphoton resonance acquire additional fine structure and
split into several closely spaced side peaks which could be observed for modes with ultra--high quality factor. The reason for that splitting 
is that high--order nonlinearities break the special symmetry specific for purely Kerr nonlinearity. Such structure of the multiphoton resonance 
intensity peak is explained by a special dependence of the tunneling rate on quasienergy which is derived from the full master equation.

\begin{acknowledgements}
This work was supported by RFBR grants 19--02--000--87a, 18--29--20032mk, 19-32-90169, and by a grant of the Foundation for the Advancement of Theoretical Physics and Mathematics 'Basis'.
\end{acknowledgements}

\appendix
\section{The continuous limit of the quantum master equation}
\label{app:qme_grad_exp}
In this Appendix, we derive 
the Fokker--Planck equation with tunneling term 
\eqref{eq:energy_fokker_planck_w_tunn} from 
the approximate form of the quantum master equation 
\eqref{eq:lindblad_reduction}, \eqref{eq:lindblad_reduction_nondiag}. 
The first step is to express $\rho_{n}^{13}$ through 
$\rho_{n}^{11}$ and $\rho_{n}^{33}$ using 
the Eq.~\eqref{eq:lindblad_reduction_nondiag}. This should be done 
differently in the quasienergy domains 
$\epsilon_\mathrm{sep} < \epsilon < \epsilon_\mathrm{crit}$ and 
$\epsilon_\mathrm{crit} < \epsilon < \epsilon_1$.

To express $\rho_{n}^{13}$ in the domain 
$\epsilon_\mathrm{sep} < \epsilon < \epsilon_\mathrm{crit}$, 
one should perform the gradient expansion of the $\gamma$--dependent 
term in \eqref{eq:lindblad_reduction_nondiag}.
It results in the following expression:
\begin{equation}
    \label{nondiag_element_below_crit} 
    \rho_{n13} = \frac{t_n(\rho_{n}^{11} - \rho_{n}^{33})}{\epsilon_{n1} - \epsilon_{n3} - \frac{i\gamma_{n13}}{2}},
\end{equation}
where
\begin{equation}
    \gamma_{n13} = \gamma\left((a^\dagger a)_{nn}^{11} + 
        (a^\dagger a)_{nn}^{33} -
                2\sum_{n'} a_{nn'}^{11}(a_{nn'}^{33})^* \right)
\end{equation}

In the domain $\epsilon_\mathrm{crit} < \epsilon < \epsilon_1$, the only
nondiagonal element of the density matrix which should be taken into account
is the element corresponding to the transition between the pair of resonant 
states, $\rho_{n_\mathrm{res}}^{13}$, and all other can be neglected. 
Because of that, $\rho_{n_{res}}^{13}$ can be immediately expressed from 
\eqref{eq:lindblad_reduction_nondiag}:
\begin{equation}
    \label{nondiag_element_res} 
    \rho_{n_{res}}^{13} = 
    \frac{t_{n_{res}}(\rho_{n_{res}}^{11} - \rho_{n_{res}}^{33})}
    {\epsilon_{n_{res}1} - \epsilon_{n_{res}3} - 
    \frac{i\tilde\gamma_{13}}{2}},
\end{equation}
\begin{multline}
        \tilde\gamma_{13} = 
        \frac{\gamma}{2}\left((a^\dagger a)_{n_{res}n_{res}}^{11} + 
        (a^\dagger a)_{n_{res}n_{res}}^{33} \right.\\- 
        \left. 2a_{n_{res}n_{res}}^{11}(a_{n_{res}n_{res}}^{33})^*\right).
\end{multline}
Then, $\rho_{n}^{13}$ from \eqref{nondiag_element_below_crit}, \eqref{nondiag_element_res} should be substituted in \eqref{eq:lindblad_reduction}. They are 
present in the term $\pm it_n(\rho^{13}_{nn} - \rho^{31}_{nn})$. 

Now, let us focus on the part containing the diagonal elements of the density matrix
$\rho^{11}_n$ and $\rho^{33}_n$. To transform the quantum master equation in the continuous form, 
one should consider $\rho^{11}_n$ and $\rho^{33}_n$ as 
the continuous functions $P^{r}(n)$ of the index $n$. Then, in the equation for
each matrix element $\rho^{rr}_n$, the gradient expansion of 
$\rho^{rr}_{n'} \equiv P^r(n')$ should be performed: 
$P^r(n') \approx P^r(n) + (n-n')\frac{\p P^r}{\p n} + \frac{1}{2}(n-n')^2\frac{\p^2 P^r}{\p n^2} + \dots$. After truncating the expansion 
up to the second order, one gets the Fokker--Planck equation with the tunneling term \eqref{eq:energy_fokker_planck_w_tunn}. The resulting drift and diffusion  
coefficients $K_r(\epsilon)$, $D_r(\epsilon)$ and the period $T_r(\epsilon)$ can be found as the contour integrals over classical trajectories of the 
nonlinear oscillator:
\begin{equation}
    \begin{gathered}
        K_r(\epsilon) = \frac{i}{2}\oint a\,da^* - a^*\,da,\\
        D_r(\epsilon) = \frac{i}{2}\oint \frac{\p H}{\p a} da - \frac{\p H}{\p a^*} da^*,\\
        T_r(\epsilon) = \int da^*da\,  \delta(\epsilon - H(a^*, a)).\\
    \end{gathered}
\end{equation}

\section{The stationary solution of 
        the Fokker--Planck equation with a tunneling term}
\label{app:fpe_stat}
In this Appendix, we present the accurate calculation of the stationary distribution function which follows from Eq.~\eqref{eq:energy_fokker_planck_w_tunn}, where
the tunneling rate $\lambda_T(\epsilon)$ is given by Eq.~\eqref{eq:lambda_T}. First of all, let us consider the domain 
$\epsilon_{sep} < \epsilon < \epsilon_{crit}$. In this domain, 
tunneling leads to the equilibration of distribution functions in regions 1 and 3. So,
$P_1(\epsilon) \approx P_3(\epsilon)$, and the equation for stationary distribution can be obtained by 
taking the sum of the equations for $P_1$ and $P_3$:
\begin{equation}
    \label{eq:stat_eq_below_crit}
   \left[\gamma (K_1 + K_3) P_{1,3} + 
    Q (D_1 + D_3) \frac{\p P_{1,3}}{\p \epsilon}\right] = 0,
\end{equation}

Then, let us consider the domain $\epsilon_{crit} < \epsilon < \epsilon_{1}$. Due to the presence of 
the delta--like peak in $\lambda_T(\epsilon)$ at $\epsilon_{res}$, the stationary distribution
function has nonzero probability flow $J$ at quasienergies $\epsilon_{crit} < \epsilon < \epsilon_{res}$ as 
presented in Fig.~\ref{fig:one_channel}
Thus, the distribution functions $P_1$ and $P_3$ obey the equations
\begin{equation}
    \label{eq:stat_eq_above_crit}
    \gamma\left[\gamma K_{1,3} P_{1,3} + 
     Q D_{1,3} \frac{\p P_{1,3}}{\p \epsilon}\right] = 
     \begin{cases}
         \mp J, & \epsilon_{crit} < \epsilon < \epsilon_{res}\\
         0, & \epsilon > \epsilon_{res}
     \end{cases}.
\end{equation}
The flow $J$ is defined from the boundary condition 
$J = \lambda_T(\epsilon_{res})(P_1(\epsilon_{res}) - P_3(\epsilon_{res}))$
which can be obtained by integrating 
Eq.~\eqref{eq:energy_fokker_planck_w_tunn} in the vicinity of $\epsilon_{res}$.
As a result, the stationary distribution function determined from the solutions of \eqref{eq:stat_eq_below_crit} and \eqref{eq:stat_eq_above_crit} read
\begin{widetext}
    \begin{equation}
        \begin{gathered}
            P_r(\epsilon) = 
            \begin{cases}
            P(\epsilon_{sep})
            \bigexp{-\frac{\gamma}{Q}\int_{\epsilon_\mathrm{sep}}^{\epsilon}
            \frac{K_1(\epsilon')+K_3(\epsilon')}{D_1(\epsilon') + D_3(\epsilon')}d\epsilon'} & \epsilon_\mathrm{sep} < \epsilon \leq \epsilon_{crit}\\
                P(\epsilon_{crit})\bigexp{-\frac{\gamma}{Q}\int^{\epsilon}_{\epsilon_\mathrm{crit}}\frac{K_r(\epsilon')}{D_r(\epsilon')}d\epsilon'} \mp
                \frac{J}{Q}\int^{\epsilon}_{\epsilon_\mathrm{crit}}\frac{d\epsilon'}{D_r(\epsilon')} \bigexp{-\frac{\gamma}{Q}\int_{\epsilon'}^{\epsilon}\frac{K_r(\tilde{\epsilon})}{D_r(\tilde{\epsilon})}d\tilde{\epsilon}}
                & \epsilon_\mathrm{crit} < \epsilon \leq \epsilon_\mathrm{res}, \quad r = 1,3\\
                P_r(\epsilon_\mathrm{res})\bigexp{-\frac{\gamma}{Q}\int_{\epsilon_\mathrm{crit}}^{\epsilon}\frac{K_r(\epsilon')}{D_r(\epsilon')}d\epsilon'} & 
        \epsilon_\mathrm{res} < \epsilon < \epsilon_1, \quad r = 1,3
            \end{cases}\\
        \end{gathered}
    \end{equation}
    \begin{equation}
        J = \frac{\tilde\gamma_{13}t^2(\epsilon_{res})P(\epsilon_{crit})
        \left(\bigexp{-\frac{\gamma}{Q}\int^{\epsilon_{res}}_{\epsilon_\mathrm{crit}}
        \frac{K_1(\tilde{\epsilon})}{D_1(\tilde{\epsilon})}d\tilde{\epsilon}} - 
        \bigexp{-\frac{\gamma}{Q}\int^{\epsilon_{res}}_{\epsilon_{crit}}
    \frac{K_3(\tilde{\epsilon})}{D_3(\tilde{\epsilon})}d\tilde{\epsilon}}
        \right)}
            {\delta\epsilon_{13\mathrm{res}}^2 + 
            \frac{\tilde\gamma_{13}^2}{4} + 
            \frac{\tilde\gamma_{13}t^2(\epsilon_{res})}{Q}
        \left(\int^{\epsilon_{res}}_{\epsilon_\mathrm{crit}}
        \frac{d\epsilon'}{D_1(\epsilon')} \bigexp{-\frac{\gamma}{Q}\int_{\epsilon'}^{\epsilon}
        \frac{K_1(\tilde{\epsilon})}{D_1(\tilde{\epsilon})}d\tilde{\epsilon}} 
            + 
         \int^{\epsilon_{res}}_{\epsilon_\mathrm{crit}}
        \frac{d\epsilon'}{D_3(\epsilon')} \bigexp{-\frac{\gamma}{Q}\int_{\epsilon'}^{\epsilon}
        \frac{K_3(\tilde{\epsilon})}{D_3(\tilde{\epsilon})}
        d\tilde{\epsilon}} \right)}.
    \end{equation}
\end{widetext}
When the detuning $\Delta$ is close to one of the values $\delta\Delta_n$, the flow $J$ has a sharp peak, because the quasienergy difference 
$\delta\epsilon_{13\mathrm{res}}$ between the resonant pair of states passes through zero. Because of this, the probability density in the classical
region 1 drops, which results in a peak in the occupation of the classical region 2.


\begin{thebibliography}{22}%
\makeatletter
\providecommand \@ifxundefined [1]{%
 \@ifx{#1\undefined}
}%
\providecommand \@ifnum [1]{%
 \ifnum #1\expandafter \@firstoftwo
 \else \expandafter \@secondoftwo
 \fi
}%
\providecommand \@ifx [1]{%
 \ifx #1\expandafter \@firstoftwo
 \else \expandafter \@secondoftwo
 \fi
}%
\providecommand \natexlab [1]{#1}%
\providecommand \enquote  [1]{``#1''}%
\providecommand \bibnamefont  [1]{#1}%
\providecommand \bibfnamefont [1]{#1}%
\providecommand \citenamefont [1]{#1}%
\providecommand \href@noop [0]{\@secondoftwo}%
\providecommand \href [0]{\begingroup \@sanitize@url \@href}%
\providecommand \@href[1]{\@@startlink{#1}\@@href}%
\providecommand \@@href[1]{\endgroup#1\@@endlink}%
\providecommand \@sanitize@url [0]{\catcode `\\12\catcode `\$12\catcode
  `\&12\catcode `\#12\catcode `\^12\catcode `\_12\catcode `\%12\relax}%
\providecommand \@@startlink[1]{}%
\providecommand \@@endlink[0]{}%
\providecommand \url  [0]{\begingroup\@sanitize@url \@url }%
\providecommand \@url [1]{\endgroup\@href {#1}{\urlprefix }}%
\providecommand \urlprefix  [0]{URL }%
\providecommand \Eprint [0]{\href }%
\providecommand \doibase [0]{http://dx.doi.org/}%
\providecommand \selectlanguage [0]{\@gobble}%
\providecommand \bibinfo  [0]{\@secondoftwo}%
\providecommand \bibfield  [0]{\@secondoftwo}%
\providecommand \translation [1]{[#1]}%
\providecommand \BibitemOpen [0]{}%
\providecommand \bibitemStop [0]{}%
\providecommand \bibitemNoStop [0]{.\EOS\space}%
\providecommand \EOS [0]{\spacefactor3000\relax}%
\providecommand \BibitemShut  [1]{\csname bibitem#1\endcsname}%
\let\auto@bib@innerbib\@empty
\bibitem [{\citenamefont {Azadpour}\ and\ \citenamefont
  {Bahari}(2019)}]{Azadpour2019}%
  \BibitemOpen
  \bibfield  {author} {\bibinfo {author} {\bibfnamefont {F.}~\bibnamefont
  {Azadpour}}\ and\ \bibinfo {author} {\bibfnamefont {A.}~\bibnamefont
  {Bahari}},\ }\href {\doibase https://doi.org/10.1016/j.optcom.2018.12.076}
  {\bibfield  {journal} {\bibinfo  {journal} {Optics Communications}\ }\textbf
  {\bibinfo {volume} {437}},\ \bibinfo {pages} {297 } (\bibinfo {year}
  {2019})}\BibitemShut {NoStop}%
\bibitem [{\citenamefont {Li}\ \emph {et~al.}(2017)\citenamefont {Li},
  \citenamefont {Ge}, \citenamefont {Wang}, \citenamefont {Martín},\ and\
  \citenamefont {Yu}}]{Li2017a}%
  \BibitemOpen
  \bibfield  {author} {\bibinfo {author} {\bibfnamefont {S.}~\bibnamefont
  {Li}}, \bibinfo {author} {\bibfnamefont {Q.}~\bibnamefont {Ge}}, \bibinfo
  {author} {\bibfnamefont {Z.}~\bibnamefont {Wang}}, \bibinfo {author}
  {\bibfnamefont {J.~C.}\ \bibnamefont {Martín}}, \ and\ \bibinfo {author}
  {\bibfnamefont {B.}~\bibnamefont {Yu}},\ }\href {\doibase
  10.1038/s41598-017-09570-x} {\bibfield  {journal} {\bibinfo  {journal}
  {Scientific Reports}\ }\textbf {\bibinfo {volume} {7}},\ \bibinfo {pages} {1}
  (\bibinfo {year} {2017})}\BibitemShut {NoStop}%
\bibitem [{\citenamefont {Pistolesi}(2018)}]{Pistolesi2018}%
  \BibitemOpen
  \bibfield  {author} {\bibinfo {author} {\bibfnamefont {F.}~\bibnamefont
  {Pistolesi}},\ }\href {\doibase 10.1103/PhysRevA.97.063833} {\bibfield
  {journal} {\bibinfo  {journal} {Phys. Rev. A}\ }\textbf {\bibinfo {volume}
  {97}},\ \bibinfo {pages} {063833} (\bibinfo {year} {2018})}\BibitemShut
  {NoStop}%
\bibitem [{\citenamefont {Gothe}\ \emph {et~al.}(2019)\citenamefont {Gothe},
  \citenamefont {Valenzuela}, \citenamefont {Cristiani},\ and\ \citenamefont
  {Eschner}}]{Gothe2019}%
  \BibitemOpen
  \bibfield  {author} {\bibinfo {author} {\bibfnamefont {H.}~\bibnamefont
  {Gothe}}, \bibinfo {author} {\bibfnamefont {T.}~\bibnamefont {Valenzuela}},
  \bibinfo {author} {\bibfnamefont {M.}~\bibnamefont {Cristiani}}, \ and\
  \bibinfo {author} {\bibfnamefont {J.}~\bibnamefont {Eschner}},\ }\href
  {\doibase 10.1103/PhysRevA.99.013849} {\bibfield  {journal} {\bibinfo
  {journal} {Phys. Rev. A}\ }\textbf {\bibinfo {volume} {99}},\ \bibinfo
  {pages} {013849} (\bibinfo {year} {2019})}\BibitemShut {NoStop}%
\bibitem [{\citenamefont {Wang}\ \emph {et~al.}(2018)\citenamefont {Wang},
  \citenamefont {Zhang}, \citenamefont {Zhang}, \citenamefont {Li},
  \citenamefont {Hu},\ and\ \citenamefont {You}}]{Wang2018}%
  \BibitemOpen
  \bibfield  {author} {\bibinfo {author} {\bibfnamefont {Y.-P.}\ \bibnamefont
  {Wang}}, \bibinfo {author} {\bibfnamefont {G.-Q.}\ \bibnamefont {Zhang}},
  \bibinfo {author} {\bibfnamefont {D.}~\bibnamefont {Zhang}}, \bibinfo
  {author} {\bibfnamefont {T.-F.}\ \bibnamefont {Li}}, \bibinfo {author}
  {\bibfnamefont {C.-M.}\ \bibnamefont {Hu}}, \ and\ \bibinfo {author}
  {\bibfnamefont {J.~Q.}\ \bibnamefont {You}},\ }\href {\doibase
  10.1103/PhysRevLett.120.057202} {\bibfield  {journal} {\bibinfo  {journal}
  {Phys. Rev. Lett.}\ }\textbf {\bibinfo {volume} {120}},\ \bibinfo {pages}
  {057202} (\bibinfo {year} {2018})}\BibitemShut {NoStop}%
\bibitem [{\citenamefont {Wang}\ \emph {et~al.}(2019)\citenamefont {Wang},
  \citenamefont {Pechal}, \citenamefont {Wollack}, \citenamefont
  {Arrangoiz-Arriola}, \citenamefont {Gao}, \citenamefont {Lee},\ and\
  \citenamefont {Safavi-Naeini}}]{Wang2019}%
  \BibitemOpen
  \bibfield  {author} {\bibinfo {author} {\bibfnamefont {Z.}~\bibnamefont
  {Wang}}, \bibinfo {author} {\bibfnamefont {M.}~\bibnamefont {Pechal}},
  \bibinfo {author} {\bibfnamefont {E.~A.}\ \bibnamefont {Wollack}}, \bibinfo
  {author} {\bibfnamefont {P.}~\bibnamefont {Arrangoiz-Arriola}}, \bibinfo
  {author} {\bibfnamefont {M.}~\bibnamefont {Gao}}, \bibinfo {author}
  {\bibfnamefont {N.~R.}\ \bibnamefont {Lee}}, \ and\ \bibinfo {author}
  {\bibfnamefont {A.~H.}\ \bibnamefont {Safavi-Naeini}},\ }\href {\doibase
  10.1103/PhysRevX.9.021049} {\bibfield  {journal} {\bibinfo  {journal} {Phys.
  Rev. X}\ }\textbf {\bibinfo {volume} {9}},\ \bibinfo {pages} {021049}
  (\bibinfo {year} {2019})}\BibitemShut {NoStop}%
\bibitem [{\citenamefont {Winkel}\ \emph {et~al.}(2020)\citenamefont {Winkel},
  \citenamefont {Borisov}, \citenamefont {Gr\"unhaupt}, \citenamefont {Rieger},
  \citenamefont {Spiecker}, \citenamefont {Valenti}, \citenamefont {Ustinov},
  \citenamefont {Wernsdorfer},\ and\ \citenamefont {Pop}}]{Winkel2020}%
  \BibitemOpen
  \bibfield  {author} {\bibinfo {author} {\bibfnamefont {P.}~\bibnamefont
  {Winkel}}, \bibinfo {author} {\bibfnamefont {K.}~\bibnamefont {Borisov}},
  \bibinfo {author} {\bibfnamefont {L.}~\bibnamefont {Gr\"unhaupt}}, \bibinfo
  {author} {\bibfnamefont {D.}~\bibnamefont {Rieger}}, \bibinfo {author}
  {\bibfnamefont {M.}~\bibnamefont {Spiecker}}, \bibinfo {author}
  {\bibfnamefont {F.}~\bibnamefont {Valenti}}, \bibinfo {author} {\bibfnamefont
  {A.~V.}\ \bibnamefont {Ustinov}}, \bibinfo {author} {\bibfnamefont
  {W.}~\bibnamefont {Wernsdorfer}}, \ and\ \bibinfo {author} {\bibfnamefont
  {I.~M.}\ \bibnamefont {Pop}},\ }\href {\doibase 10.1103/PhysRevX.10.031032}
  {\bibfield  {journal} {\bibinfo  {journal} {Phys. Rev. X}\ }\textbf {\bibinfo
  {volume} {10}},\ \bibinfo {pages} {031032} (\bibinfo {year}
  {2020})}\BibitemShut {NoStop}%
\bibitem [{\citenamefont {Muppalla}\ \emph {et~al.}(2018)\citenamefont
  {Muppalla}, \citenamefont {Gargiulo}, \citenamefont {Mirzaei}, \citenamefont
  {Venkatesh}, \citenamefont {Juan}, \citenamefont {Grunhaupt}, \citenamefont
  {Pop},\ and\ \citenamefont {Kirchmair}}]{Muppalla2018}%
  \BibitemOpen
  \bibfield  {author} {\bibinfo {author} {\bibfnamefont {P.~R.}\ \bibnamefont
  {Muppalla}}, \bibinfo {author} {\bibfnamefont {O.}~\bibnamefont {Gargiulo}},
  \bibinfo {author} {\bibfnamefont {S.~I.}\ \bibnamefont {Mirzaei}}, \bibinfo
  {author} {\bibfnamefont {B.~P.}\ \bibnamefont {Venkatesh}}, \bibinfo {author}
  {\bibfnamefont {M.~L.}\ \bibnamefont {Juan}}, \bibinfo {author}
  {\bibfnamefont {L.}~\bibnamefont {Grunhaupt}}, \bibinfo {author}
  {\bibfnamefont {I.~M.}\ \bibnamefont {Pop}}, \ and\ \bibinfo {author}
  {\bibfnamefont {G.}~\bibnamefont {Kirchmair}},\ }\href@noop {} {\bibfield
  {journal} {\bibinfo  {journal} {Phys. Rev. B}\ }\textbf {\bibinfo {volume}
  {97}},\ \bibinfo {pages} {024518} (\bibinfo {year} {2018})}\BibitemShut
  {NoStop}%
\bibitem [{\citenamefont {Maslova}\ \emph
  {et~al.}(2019{\natexlab{a}})\citenamefont {Maslova}, \citenamefont
  {Mantsevich}, \citenamefont {Arseyev},\ and\ \citenamefont
  {Sokolov}}]{PhysRevB.100.035307}%
  \BibitemOpen
  \bibfield  {author} {\bibinfo {author} {\bibfnamefont {N.~S.}\ \bibnamefont
  {Maslova}}, \bibinfo {author} {\bibfnamefont {V.~N.}\ \bibnamefont
  {Mantsevich}}, \bibinfo {author} {\bibfnamefont {P.~I.}\ \bibnamefont
  {Arseyev}}, \ and\ \bibinfo {author} {\bibfnamefont {I.~M.}\ \bibnamefont
  {Sokolov}},\ }\href {\doibase 10.1103/PhysRevB.100.035307} {\bibfield
  {journal} {\bibinfo  {journal} {Phys. Rev. B}\ }\textbf {\bibinfo {volume}
  {100}},\ \bibinfo {pages} {035307} (\bibinfo {year}
  {2019}{\natexlab{a}})}\BibitemShut {NoStop}%
\bibitem [{\citenamefont {Gibbs}\ \emph {et~al.}(1976)\citenamefont {Gibbs},
  \citenamefont {McCall},\ and\ \citenamefont {Venkatesan}}]{Gibbs1976}%
  \BibitemOpen
  \bibfield  {author} {\bibinfo {author} {\bibfnamefont {H.~M.}\ \bibnamefont
  {Gibbs}}, \bibinfo {author} {\bibfnamefont {S.~L.}\ \bibnamefont {McCall}}, \
  and\ \bibinfo {author} {\bibfnamefont {T.~N.~C.}\ \bibnamefont
  {Venkatesan}},\ }\href {\doibase 10.1103/physrevlett.36.1135} {\bibfield
  {journal} {\bibinfo  {journal} {Phys. Rev. Lett}\ }\textbf {\bibinfo {volume}
  {36}},\ \bibinfo {pages} {1135} (\bibinfo {year} {1976})}\BibitemShut
  {NoStop}%
\bibitem [{\citenamefont {Bonifacio}\ and\ \citenamefont
  {Lugiato}(1978)}]{Bonifacio1978}%
  \BibitemOpen
  \bibfield  {author} {\bibinfo {author} {\bibfnamefont {R.}~\bibnamefont
  {Bonifacio}}\ and\ \bibinfo {author} {\bibfnamefont {L.~A.}\ \bibnamefont
  {Lugiato}},\ }\href {\doibase 10.1103/PhysRevA.18.1129} {\bibfield  {journal}
  {\bibinfo  {journal} {Phys. Rev. A}\ }\textbf {\bibinfo {volume} {18}},\
  \bibinfo {pages} {1129} (\bibinfo {year} {1978})}\BibitemShut {NoStop}%
\bibitem [{\citenamefont {Maslova}\ \emph
  {et~al.}(2019{\natexlab{b}})\citenamefont {Maslova}, \citenamefont {Anikin},
  \citenamefont {Gippius},\ and\ \citenamefont {Sokolov}}]{Maslova2019}%
  \BibitemOpen
  \bibfield  {author} {\bibinfo {author} {\bibfnamefont {N.~S.}\ \bibnamefont
  {Maslova}}, \bibinfo {author} {\bibfnamefont {E.~V.}\ \bibnamefont {Anikin}},
  \bibinfo {author} {\bibfnamefont {N.~A.}\ \bibnamefont {Gippius}}, \ and\
  \bibinfo {author} {\bibfnamefont {I.~M.}\ \bibnamefont {Sokolov}},\ }\href
  {\doibase 10.1103/PhysRevA.99.043802} {\bibfield  {journal} {\bibinfo
  {journal} {Phys. Rev. A}\ }\textbf {\bibinfo {volume} {99}},\ \bibinfo
  {pages} {043802} (\bibinfo {year} {2019}{\natexlab{b}})}\BibitemShut
  {NoStop}%
\bibitem [{\citenamefont {Anikin}\ \emph {et~al.}(2019)\citenamefont {Anikin},
  \citenamefont {Maslova}, \citenamefont {Gippius},\ and\ \citenamefont
  {Sokolov}}]{Anikin2019}%
  \BibitemOpen
  \bibfield  {author} {\bibinfo {author} {\bibfnamefont {E.~V.}\ \bibnamefont
  {Anikin}}, \bibinfo {author} {\bibfnamefont {N.~S.}\ \bibnamefont {Maslova}},
  \bibinfo {author} {\bibfnamefont {N.~A.}\ \bibnamefont {Gippius}}, \ and\
  \bibinfo {author} {\bibfnamefont {I.~M.}\ \bibnamefont {Sokolov}},\ }\href
  {\doibase 10.1103/PhysRevA.100.043842} {\bibfield  {journal} {\bibinfo
  {journal} {Phys. Rev. A}\ }\textbf {\bibinfo {volume} {100}},\ \bibinfo
  {pages} {043842} (\bibinfo {year} {2019})}\BibitemShut {NoStop}%
\bibitem [{\citenamefont {Keldysh}(1965)}]{Keldysh1965}%
  \BibitemOpen
  \bibfield  {author} {\bibinfo {author} {\bibfnamefont {L.~V.}\ \bibnamefont
  {Keldysh}},\ }\href@noop {} {\bibfield  {journal} {\bibinfo  {journal} {Sov.
  Phys. JETP}\ }\textbf {\bibinfo {volume} {20}},\ \bibinfo {pages} {1307}
  (\bibinfo {year} {1965})}\BibitemShut {NoStop}%
\bibitem [{\citenamefont {Drummond}\ and\ \citenamefont
  {Walls}(1980)}]{Drummond1980}%
  \BibitemOpen
  \bibfield  {author} {\bibinfo {author} {\bibfnamefont {P.~D.}\ \bibnamefont
  {Drummond}}\ and\ \bibinfo {author} {\bibfnamefont {D.~F.}\ \bibnamefont
  {Walls}},\ }\href {\doibase 10.1088/0305-4470/13/2/034} {\bibfield  {journal}
  {\bibinfo  {journal} {J. of Physics A: Mathematical and General}\ }\textbf
  {\bibinfo {volume} {13}},\ \bibinfo {pages} {725} (\bibinfo {year}
  {1980})}\BibitemShut {NoStop}%
\bibitem [{\citenamefont {Risken}\ \emph {et~al.}(1987)\citenamefont {Risken},
  \citenamefont {Savage}, \citenamefont {Haake},\ and\ \citenamefont
  {Walls}}]{Risken1987}%
  \BibitemOpen
  \bibfield  {author} {\bibinfo {author} {\bibfnamefont {H.}~\bibnamefont
  {Risken}}, \bibinfo {author} {\bibfnamefont {C.}~\bibnamefont {Savage}},
  \bibinfo {author} {\bibfnamefont {F.}~\bibnamefont {Haake}}, \ and\ \bibinfo
  {author} {\bibfnamefont {D.~F.}\ \bibnamefont {Walls}},\ }\href {\doibase
  10.1103/physreva.35.1729} {\bibfield  {journal} {\bibinfo  {journal} {Phys.
  Rev. A}\ }\textbf {\bibinfo {volume} {35}},\ \bibinfo {pages} {1729}
  (\bibinfo {year} {1987})}\BibitemShut {NoStop}%
\bibitem [{\citenamefont {Haken}(1965)}]{Haken1965}%
  \BibitemOpen
  \bibfield  {author} {\bibinfo {author} {\bibfnamefont {H.}~\bibnamefont
  {Haken}},\ }\href@noop {} {\bibfield  {journal} {\bibinfo  {journal}
  {Zeitschrift für Physik}\ }\textbf {\bibinfo {volume} {219}},\ \bibinfo
  {pages} {411} (\bibinfo {year} {1965})}\BibitemShut {NoStop}%
\bibitem [{\citenamefont {Risken}(1965)}]{Risken1965}%
  \BibitemOpen
  \bibfield  {author} {\bibinfo {author} {\bibfnamefont {H.}~\bibnamefont
  {Risken}},\ }\href@noop {} {\bibfield  {journal} {\bibinfo  {journal}
  {Zeitschrift für Physik}\ }\textbf {\bibinfo {volume} {186}},\ \bibinfo
  {pages} {85} (\bibinfo {year} {1965})}\BibitemShut {NoStop}%
\bibitem [{\citenamefont {Graham}\ and\ \citenamefont
  {Haken}(1970)}]{Graham1970}%
  \BibitemOpen
  \bibfield  {author} {\bibinfo {author} {\bibfnamefont {R.}~\bibnamefont
  {Graham}}\ and\ \bibinfo {author} {\bibfnamefont {H.}~\bibnamefont {Haken}},\
  }\href@noop {} {\bibfield  {journal} {\bibinfo  {journal} {Zeitschrift für
  Physik}\ }\textbf {\bibinfo {volume} {219}},\ \bibinfo {pages} {246}
  (\bibinfo {year} {1970})}\BibitemShut {NoStop}%
\bibitem [{\citenamefont {{Vogel}}\ and\ \citenamefont
  {{Risken}}(1988)}]{Vogel1988}%
  \BibitemOpen
  \bibfield  {author} {\bibinfo {author} {\bibfnamefont {K.}~\bibnamefont
  {{Vogel}}}\ and\ \bibinfo {author} {\bibfnamefont {H.}~\bibnamefont
  {{Risken}}},\ }\href {\doibase 10.1103/PhysRevA.38.2409} {\bibfield
  {journal} {\bibinfo  {journal} {\pra}\ }\textbf {\bibinfo {volume} {38}},\
  \bibinfo {pages} {2409} (\bibinfo {year} {1988})}\BibitemShut {NoStop}%
\bibitem [{\citenamefont {Maslova}\ \emph {et~al.}(2007)\citenamefont
  {Maslova}, \citenamefont {Johne},\ and\ \citenamefont
  {Gippius}}]{Maslova2007}%
  \BibitemOpen
  \bibfield  {author} {\bibinfo {author} {\bibfnamefont {N.~S.}\ \bibnamefont
  {Maslova}}, \bibinfo {author} {\bibfnamefont {R.}~\bibnamefont {Johne}}, \
  and\ \bibinfo {author} {\bibfnamefont {N.~A.}\ \bibnamefont {Gippius}},\
  }\href {\doibase 10.1134/S0021364007140123} {\bibfield  {journal} {\bibinfo
  {journal} {JETP Letters}\ }\textbf {\bibinfo {volume} {86}},\ \bibinfo
  {pages} {126} (\bibinfo {year} {2007})}\BibitemShut {NoStop}%
\bibitem [{\citenamefont {Vogel}\ and\ \citenamefont
  {Risken}(1990)}]{Vogel1990}%
  \BibitemOpen
  \bibfield  {author} {\bibinfo {author} {\bibfnamefont {K.}~\bibnamefont
  {Vogel}}\ and\ \bibinfo {author} {\bibfnamefont {H.}~\bibnamefont {Risken}},\
  }\href {\doibase 10.1103/PhysRevA.42.627} {\bibfield  {journal} {\bibinfo
  {journal} {\pra}\ }\textbf {\bibinfo {volume} {42}},\ \bibinfo {pages} {627}
  (\bibinfo {year} {1990})}\BibitemShut {NoStop}%
\end{thebibliography}
\end{document}